\documentclass[12pt]{article} \usepackage{latexsym}

\def\cS{{\cal S}}
\def\cH{{\cal H}}
\def\ket#1{\mid~\!\!\!{#1}~\!\!\rangle}
\def\bra#1{\langle~\!\!{#1}~\!\!\!\mid}
\def\+-{\buildrel + \over -}
\def\IF {if and only if }
\def\QM{quantum mechanics }
\def\qm{quantum mechanics}
\def\Q{quantum }

\def\ON{orthonormal }
\def\${\enskip$}
\def\M{measurement }
\def\m{measurement}
\def\WF{wavefunction }

\def\N{_{12\dots N}}
\def\RFP{relative-frequency procedure }

\def\E{ensemble }

\def\EX{experimental }

\def\RF{relative frequency }
\def\rf{relative frequency}
\def\P{probability }

\def\T{theoretical }

\def\EM{ensemble-measurement }

\begin{document}

\begin{center}

{\large \bf
INDETERMINATE PTOBABILITIES
AND THE WEAK QUANTUM LAW\\ OF LARGE NUMBERS}
\vspace{.3cm}

\bf \large
 Fedor Herbut\\
\end{center}

\noindent {\footnotesize \it Serbian Academy of
Sciences and Arts, Serbia,
Belgrade, Knez Mihajlova 35}

\vspace{0.3cm} \noindent \rule{13.6cm}{.4pt}

{\bf \large ABSTRACT} The  \Q probabilistic convergence in \m , distinct from mathematical convergence, is derived for indeterminate probabilities \$0<\bra{\psi}P\ket{\psi}<1\$, \$P\$ being a projector (event, property), from the weak \Q law of large numbers. This is presented in three theorems. The first establishes the necessary bridge between ensemble theory and experiment.
The second  analyzes the most important theoretical ensemble entity: the eigen-projector of the relative frequency operator. Its physical meaning is the experimental relative frequency. The third theorem formulates the \Q probabilistic convergence, which is the final result of this investigation.

\noindent \rule{13.6cm}{.4pt}

\rm

%1
%%%%%%%%%%%%%%%%%%%%%%%%%%%%%%%%%%%%%%%%%%%%%%%%%%%%%%
%%%%%%%%%%%%%%%%%%%%%%%%%%%%%%%%%%%%%%%%%%%%%%%%%%%%%%
%%%%%%%%%%%%%%%%%%%%%%%%%%%%%%%%%%%%%%%%%%%%%%%%%%%%%%
\section{Introduction}
%%%%%%%%%%%%%%%%%%%%%%%%%%%%%%%%%%%%%%%%%%%%%%%%%%%%%%
%%%%%%%%%%%%%%%%%%%%%%%%%%%%%%%%%%%%%%%%%%%%%%%%%%%%%%
%%%%%%%%%%%%%%%%%%%%%%%%%%%%%%%%%%%%%%%%%%%%%%%%%%%%%%

After the recent {\it \$\Psi$-ontic} breakthrough
\cite{Pusey}, \cite{Colbeck}, \cite{Hardy}, \cite{Leifer} the question {\it how is reality contained}
in such an intricate concept as the wavefunction
becomes more actual than ever.

The first thing that comes to mind is the {\it e-e
link} (eigenvalue-eigenstate link). As it is nicely
explained in \cite{Leifer} (subsection 3.2 there), it is a tenet that says that any observable
(Hermitian operator) \${\bar M}\$ in a given wave-
function \$\psi$ has the value \$\bar m\$ as its ontic property (property of the ontic state or,
equivalently, property of each individual system in
the \Q state \$\ket{\psi}$) \IF the eigenvalue equation
\${\bar M}\ket{\psi} =\bar m\ket{\psi}\$ is valid.

In this article we assume that a single-particle projector \$P\$ (event, property) and a wave function \$\ket{\psi}\$ are  given and we {\it investigate the \M } of the probability \$\bra{\psi}P\ket{\psi}\$. As it is well known, \$0\leq\bra{\psi}P\ket{\psi}\leq 1\$ is always valid. The extreme value \$\bra{\psi}P\ket{\psi}=1\$
is equivalent to \$P\ket{\psi}=\ket{\psi}\$\\
\vspace{0.1cm}
\noindent \rule{5cm}{.4pt}\\
fedorh@sanu.ac.rs\\
(cf Loose end 2 in the Appendix). Hence, it belongs to the e-e link. The other extreme, \$\bra{\psi}P\ket{\psi}=0\$
is equivalent to \$P^{\perp}\ket{\psi}=0\$, where \$P^{
\perp}\equiv I-P\$, \$I\$ being the identity operator (cf {\it ibid.}) Thus, also this case belongs to the e-e link.

The  e-e link deserves a separate detailed analysis. It will be done elsewhere.\\

One cannot help wondering what about the reality of the other observables \$M\$, for which \$\ket{\psi}\$ is not an eigen-state. It is an almost universally accepted empirical fact
that the \WF describes the corresponding ensembles
very well. Let us outline how this description goes.
The {\it relative-frequency procedure} for the \M of
the probability
of an event (projector) \$P\$ is well known: One performs complete \M of the event (making it
occur or not) on N identically prepared
non-interacting copies of the \Q system in the same
given \Q state \$\ket{\psi}\$. The N systems are said to make up an {\it
ensemble}. If the event
\$P\$ occurs on K systems \$(0\leq
K\leq N)\$, then  K/N, this {\it experimental  relative
frequency}, is taken to be approximately
equal to the probability of the event (or property) \$P\$ in the state \$\ket{\psi}\$.

The \RFP is often believed to be utilized {\it on empirical
grounds}.  The simplest view is  to take mathematical convergence to be its mathematically precise form: that the experimental relative frequencies, if taken in an infinite sequence, have  the probability of \$P\$ in \$\psi\$ as their limit value. This can be supported by Kolmogorovian arguments, e. g. \cite{Lahti}.

The {\it aim} of this article is to clarify if this is true. If it is not, one must find its precise modification. And all this should follow necessarily from the \Q formalism (by which I mean \QM minus its interpretation), which  is thoroughly verified by innumerable experiments and hence can be thought to contain the \Q laws of nature.

No probability theory can determine to what objects it is applicable. But it {\it should determine how it is applied}. It is shown in this article that one can derive an answer along these lines from the weak \Q law of large numbers.\\

"Projector" and "event", as well as "Hermitian
operator" and "observable" are used interchangeably
throughout. Analogously, "norm-one vector", "pure state" and "wavefunction" are not distinguished, as well as "density operator" and "general state".\\

The article {\it organization} can be seen in the titles of the sections and subsections.

Section 2. Ensembles and the relative-frequency operator

Section 3. Bridging the gap between theory and experiment

Section 4. Probabilities of experimental relative frequencies

Section 5. Quantum probabilistic convergence

Section 6. Concluding remarks (A.,B.,C.,D.,E.)

Appendix. Two Loose ends are tied up.\\

The content of Section 2 is more or less known . (It is scattered in the literature.) The rest is hopefully original.\\

%2
%%%%%%%%%%%%%%%%%%%%%%%%%%%%%%%%%%%%%%%%%%%%%%%%%%%%%%%%%%%%%%%%%%%%%%%%%%%%%
%%%%%%%%%%%%%%%%%%%%%%%%%%%%%%%%%%%%%%%%%%%%%%%%%%%%%%%%%%%%%%%%%%%%%%%%%%%%%
%%%%%%%%%%%%%%%%%%%%%%%%%%%%%%%%%%%%%%%%%%%%%%%%%%%%%%%%%%%%%%%%%%%%%%%%%%%%%
\section{Ensembles and the\\ Relative-Frequency Operator}
%%%%%%%%%%%%%%%%%%%%%%%%%%%%%%%%%%%%%%%%%%%%%%%%%%%%%%%%%%%%%%%%%%%%%%%%%%%%%
%%%%%%%%%%%%%%%%%%%%%%%%%%%%%%%%%%%%%%%%%%%%%%%%%%%%%%%%%%%%%%%%%%%%%%%%%%%%%
%%%%%%%%%%%%%%%%%%%%%%%%%%%%%%%%%%%%%%%%%%%%%%%%%%%%%%%%%%%%%%%%%%%%%%%%%%%%%

Classically each event always either occurs or its opposite event occurs. Quantum-mechanically, one performs a \M of an
event \$P\$ as an observable, the spectral form of which is \$P=1\times P+0\times
P^{\perp}\$, where \$P^{\perp}\equiv I-P\$ (\$I\$
being the identity operator) is the opposite event (the ortho-complementary projector). If
\$\ket{\psi }\$ is a given \WF , then the \Q formalism
yields the well-known probability \$0\leq\bra{\psi
}P\ket{\psi }\leq 1\$. If \$P\$ is an eigen-projector of an observable \$M\$ to which the e-e link does not apply (see the Introduction), then the probability has a value in the open
interval.

In a single complete \M of \$P\$ in \$\ket{\psi}\$ the eigenvalue 1 or 0 is
obtained with no regard to the value of the
probability. It is the general statistical procedure that one takes resort to ensembles. Then the
relevance of the so-called relative-frequency
operator emerges in the \Q formalism.\\

The  {\it relative-frequency operator} \$F^{P}\N\$ is associated with a given  projector \$P\$ that acts in the
single-system state space \$\cH\$. The operator
\$F^{P}\N\$ is defined in the \$N$-system state space
\$\cH\N\equiv\cH_1\otimes\cH_2\otimes\dots\otimes\cH_N\$,
where the tensor factor spaces \$\cH_n,\enskip n=1,2,\dots
,N\$ are isomorphic to \$\cH\$, and the operators in
them (like \$P_1,P_2,\dots ,P_N\$) are equivalent to
the equally denoted operator \$P\$  in \$\cH\$. The
definition of \$F\N^{P}\$ goes as follows:

$$F\N^{P}\equiv (1/N)\sum_{n=1}^N P^n\N,\eqno{(1a)}$$

$$P^n\N\equiv (I_1\otimes I_2\otimes\dots\otimes
I_{n-1}
\otimes P_n\otimes I_{n+1}\otimes\dots\otimes
I_N),\quad n=1,2,\dots ,N ,\eqno{(1b)}$$
where \$I_n\$ is the identity operator in \$\cH_n\$.
For \$n=1\$ and \$n=N\$ slight modification
in (1b) is required, which makes \$P_1 $ the first factor or \$P_N\$ the last factor respectively.\\

{\it LEMMA 1}. The {\it spectrum} of
\$F_{\N}^{P}\$ is \$\{0,1/N,2/N,\dots
,(N-1)/N,1\}\$.\\

{\it PROOF} Let \$\{\ket{u=1,q^u}:\forall
q^u\}\$ and \$\{\ket{u=0,q^u}:\forall q^u\}\$ together be a complete eigen-basis of \$P\$.
The index \$q^u\$ enumerates the possible multiplicity (degeneracy) of the eigenvalue \$u\$ of \$P\$.

Then in each term of
$$F_{\N}^{P}
\Big(\ket{u_1,q^{u_1}}_1\ket{u_2,q^{u_2}}_2\dots
\ket{u_N,q^{u_N}}_N\Big)=$$  $$(1/N)\sum_{n=1}^N
P\N^n\Big(\ket{u_1,q^{u_1}}_1\ket{u_2,q^{u_2}}_2\dots
\ket{u_N,q^{u_N}}_N\Big),\eqno{(2)}$$
when the operator string in (1b) is applied to the vector string, each factor operator acting on the corresponding factor state, then the crucial factor among the resulting vectors is \$P_n\ket{u_n,q^{u_n}}=
\delta_{u_n,1}\ket{u_n,q^{u_n}}\$.

Let there be K single-system states in the strung of states that are 1-eigen-states of \$P\$. Then we omit the zero terms (in which \$P\$ acts on its 0-eigen-state) in the sum.  \$K\$ terms remain unchanged. Hence
$$F_{\N}^{P}
\Big(\ket{u_1,q^{u_1}_1}_1\ket{u_2,q^{u_2}_2}_2\dots
\ket{u_N,q^{u_N}_N}_N\Big)=$$  $$
(K/N)
\Big(\ket{u_1,q^{u_1}_1}_1\ket{u_2,q^{u_2}_2}_2\dots
\ket{u_N,q^{u_N}_N}_N\Big).\eqno{(3)}$$
When all strings of eigen-vectors are taken into account on the LHS, then it follows that the claimed eigenvalues \$K/N\$
appear for all \$K=0,1,\dots ,N\$.\hfill
$\Box$.\\

{\it    DEFINITION 1} We call the numbers
\$\{K_N/N:K_N=0,1,\dots ,N\}\$ {\it \T relative
frequencies}.\\

One should note the conceptual distinction that the theoretical relative frequencies are
not the same as the \E relative frequencies. The
former are entities of the \Q formalism. The latter
are entities observed in the laboratory. The first task of this study is to understand the relation
between the \T relative frequencies and the \EX ones. This will be done in the next section.\\

Let us write the unique {\it spectral form}  of
\$F_{\N}^{P}\$ , i. e.,
the one with no repetition in the eigenvalues, as
$$F_{\N}^{P}=\sum_{K=0}^N(K/N) Q_{\N}^K,\eqno{(4a)}$$ where \$\{Q\N^K:K=0,1,\dots ,N\}\$ are the eigen-projectors. The spectral form is
accompanied by the (orthogonal projector) spectral decomposition of the identity operator \$I_{\N}\$ (also called the completeness relation)
$$I_{\N}=\sum_{K=0}^NQ_{\N}^K.\eqno{(4b)}$$\\

The manner how the given single-system projector \$P\$ determines the eigen-projectors \$\{ Q\N^K:K=0,1,2,\dots ,N;N=1,2,\dots ,\infty\}\$
will   be given below (cf relation (16)), where they will furnish useful information.\\

For any single-system state vector   \$\ket{\psi}\$, the
{\it ensemble state vector} \$\ket{\Psi}\N\$ is defined as
the tensor product, or the N-th tensor power:

$$\ket{\Psi}\equiv\prod_{n=1}^{\otimes,N}\ket{\psi}_n,
\eqno{(5)}$$
where \$\{\ket{\psi}_1,\ket{\psi}_2,\dots ,\ket{\psi}_n\}\$ are
isomorphic to \$\ket{\psi}\$.\\

{\it LEMMA 2} $$\bra{\Psi}\N
F\N^{P}\ket{\Psi}\N=p\equiv\bra{\psi} P\ket{\psi}.\eqno{(6)}$$

{\it PROOF} The definitions (1a,b) of \$F^{ P}\N\$ and (5) of \$\Psi\N\$ make the LHS of (6) a sum of strings of corresponding single-system entities. When all this is written out and evaluation factor-by-factor is performed in each term, one ends up with the sum of products of numbers. All but one in each of them are \$\bra{\psi}_{n'}\ket{\psi}_{n'}=1\$, and the exceptional one is \$\bra{\psi}_n P_n\ket{\psi}_n\$. So that finally \$LHS(6)=(1/N)Np\$ as claimed.\hfill $\Box$\\

{\it REMARK 1} Relation (6) makes the
probability \$p\$  the {\it average \T\RF} in the
ensemble state \$\ket{\Psi}\N \$ for every value of N. This becomes even
more obvious when one substitutes the spectral form
(4a) of \$\hat F\N^{P}\$ in (6):
$$p=\sum_{K=0}^N \Big(K/N\Big)\bra{\Psi}\N
 Q\N^K\ket{\Psi}\N .\eqno{(7)}$$\\

The so-called {\it variance} (square of the standard
deviation)
$$V_N=V_N(F^{P}\N,\ket{\Psi}\N,p)\equiv
\bra{\Psi}\N(F^{ P}\N-p)^2\ket{\Psi}\N\eqno{(8)}$$
plays the {\it main role} in the {\it \Q weak law of large numbers} (shortly "weak law") \cite{Farhi}.  It
characterizes the way how the eigenvalues
\$\{K/N:K=0,1,\dots ,N\}\$ of\$F^{P}\N\$
are distributed around the average value \$p\$ in the ensemble state \$\ket{\Psi}\N\$.

As it is well known, a function of an operator has the same eigen-projectors as the latter, and the eigenvalues of the former equal the given function of the eigenvalues of the latter. Thus, the spectral form (of the function of \$F^{P}\N \$)
$$(F^{P}\N-p)^2=\sum_{K=0}^N[(K/N)-p]^2\hat Q\N^K,\eqno{(9)}$$ when substituted in (8), it entails
$$V_N=\sum_{K=0}^N(K/N-p)^2\bra{\Psi
}Q\N^K\ket{\Psi }.\eqno{(10)}$$

Form (10) of \$V_N\$ will be made important use of in the sequel. But to evaluate \$V_N\$, we return to (8), and we evaluate
 $$(F^{P}\N-p)^2=(F\N^{P})^2-2pF\N^{P}+p^2.
 $$
Then
$$V_N=\bra{\Psi}\N(F^{P}\N)^2\ket{\Psi}\N
-p^2.\eqno{(11)}$$ Now we are prepared to state and prove the
following important claim.\\

{\it LEMMA 3}
$$V_N=p(1-p)/N,\quad N=1,2,\dots ,\infty.\eqno{(12)}$$

{\it PROOF} Definitions (1a,b) can be utilized having
in mind the well known fact that there are
\$N(N-1)/2\$ distinct pairs of  \$N\$
systems (much used in \$N$-system interaction).
Besides, we utilize the idempotency of the single-system projectrs.
$$\bra{\Psi}\N(F^{P}\N )^2\ket{\Psi}\N=
(1/N^2)\sum_{n=1}^N\sum_{n'=1}^N\bra{\Psi}\N ( P^n\N
P^{n'}\N )\ket{\Psi}\N=$$
$$(1/N^2)\bra{\Psi}\N\Big(\sum_{n<n'}(P^n\N P^{n'}\N
) +$$  $$ \sum_{n>n'}(P^n\N P^{n'}\N )
+\sum_{n=n'}(P^n\N P^{n'}\N ) \Big)\ket{\Psi}\N=$$
$$2(1/N^2)\Big(N(N-1)/2\Big)p^2+(1/N^2)Np=$$  $$
(N^2-N)/N^2)p^2+p/N=p^2-(p^2/N)+p/N.\eqno{(13)}$$
Then (11b) implies the claimed relation (12).\hfill
$\Box$\\

The \Q formalism allows one to rewrite the variance
in an interesting way, which is relevant to the aim
of this study. Namely, (8) can be
rewritten as

$$V_N=||F^{P}\N\ket{\Psi}\N -p\ket{\Psi}\N
||^2.\eqno{(14)}$$
Then (12) gives
$$||F^{P}\N\ket{\Psi}\N -p\ket{\Psi}\N ||^2=p(1-p)/N
.\eqno{(15)}$$

Relation (15) expresses the claim that the {\it distance} between
the two sequences \$\{F^{P}\N\ket{\Psi}\N :N=1,2,\dots
,\infty\}\$ and
\$\{p\ket{\Psi}\N :N=1,2,\dots ,\infty\}\$ goes to zero as N
goes to infinity; or, as one can say, they 'meet'
asymptotically, i. e., in the infinity. This is the
conceptual definition of the {\it \Q weak law}. (Its operational definition is given by definition (8) of \$V_N\$ and relation (12) in Lemma 3.)

Neither of the two mentioned sequences converges in the
orthogonal sum \$\sum_{N=1}^{\oplus
,\infty}\cH_{\N}\$, which is, if the one-dimensional
vacuum space is added, the well-known space of second
quantization (cf Loose end 1 in the Appendix). This
state space is as far as the experimentally supported
\Q formalism goes. (More on this in concluding remark B in subsection 9.1)\\

Between (15) and the \RFP (stated in the Introduction) there still looms a large
gap. We start to bridge this
gap in the next section.\\

%3     last (15)
%%%%%%%%%%%%%%%%%%%%%%%%%%%%%%%%%%%%%%%%%%%%%%%%%%%%%%%%%%%%%%%%%%%%%%%%%%%%%
%%%%%%%%%%%%%%%%%%%%%%%%%%%%%%%%%%%%%%%%%%%%%%%%%%%%%%%%%%%%%%%%%%%%%%%%%%%%%
%%%%%%%%%%%%%%%%%%%%%%%%%%%%%%%%%%%%%%%%%%%%%%%%%%%%%%%%%%%%%%%%%%%%%%%%%%%%%
\section{Bridging the Gap between Theory and Experiment}
%%%%%%%%%%%%%%%%%%%%%%%%%%%%%%%%%%%%%%%%%%%%%%%%%%%%%%%%%%%%%%%%%%%%%%%%%%%%%
%%%%%%%%%%%%%%%%%%%%%%%%%%%%%%%%%%%%%%%%%%%%%%%%%%%%%%%%%%%%%%%%%%%%%%%%%%%%%
%%%%%%%%%%%%%%%%%%%%%%%%%%%%%%%%%%%%%%%%%%%%%%%%%%%%%%%%%%%%%%%%%%%%%%%%%%%%%

To begin with, let us clarify the connection between the theoretical and the experimental relative frequencies \$K/N\$. To this end let us put the question how one achieves eigen-values, i. e., how one performs a complete measurement,
of the \RF operator \$F\N^{P}\$.

To obtain
an answer, let us make the mental transition from
theory to experiment, and try to
give an answer from the experimental side. As it was mentioned in the Introduction, what is
typically done, one performs a {\it complete \M of
the given single-system event \$P\$} as an observable
\$(P=1\times P+0\times P^{\perp})\$ on each of the single systems in
the given single-system state \$\psi\$ in the
ensemble of \$N\$ systems in the laboratory. One
obtains 1 (occurrence) or zero (occurrence of \$P^{\perp}\$) on
each sample. Adding up the results and dividing by N, one arrives at the {\it \EX \RF \$K/N\$}. It is an \EM result. Besides, it is a statistical notion because it concerns real systems in the laboratory.\\

For further progress, one should utilize the fact  that a certainty relation \$\bra{\psi} P\ket{\psi}=1\$ is more practically equivalently written as \$P\ket{\psi}=\ket{\psi}\$ (proof in Loose end 2). But first we need an explicit expression for \$Q\N^K\$.\\

{\it LEMMA 4} The eigen-projectors
\$Q_{\N}^K\$ of \$F_{\N}^{P}\$ are explicitly
determined by \$P\$ as follows:
$$Q_{\N}^K=P_1\otimes P_2\otimes\dots \otimes
P_K\otimes P^{\perp}_{K+1}
\otimes P^{\perp}_{K+2}\otimes\dots \otimes
P_N^{\perp}\quad \textbf{+}\quad\mbox{the rest of the terms}.\eqno{(16)}$$
"The rest of the terms" are determined by the rest of the distinct combinations without repetition of order K
out of \$\{1,2,\dots ,N\}\$, shortly combs$^K_N$,
specifying the tensor factors \$P_n\$ in the corresponding term in \$Q\N^K\$. The rest of (N-K) factors in the same term are the ortho-complementary projectors
(or opposite events)  \$P^{\perp}_n\$.\\

{\it PROOF} is obtained in terms of a complete \ON eigen-basis of \$F\N^{P}\$ consisting of
N-tensor-factor products of vectors from a complete \ON
eigen-basis of \$P\$: \$\{\ket{u=1,q^u}:\forall
q^u\}\$ and \$\{\ket{u=0,q^u}:\forall q^u\}\$
(cf proof of Lemma 1). One can write \$\ket{u,q^u}=P\ket{u,q^u}\$ if \$u=1\$ and \$\ket{u,q^u}=P^{\perp}\ket{u,q^u}\$ if \$u=0\$. This allows to rewrite the latter eigen-basis as
\$\{P\ket{u=1,q^u}:\forall
q^u\}\$ and \$P^{\perp}\ket{u=0,q^u}:\forall q^u\}\}\$, and then make the tensor multiplication into the N-factor ensemble basis vectors.

One obtains the eigenvalue 1 of
\$Q_{\N}^K\$ precisely when application of
\$F_{\N}^{P}\$ to the same N-system vector gives the
eigenvalue \$K/N\$, i. e., when there are precisely \$K\$ projectors \$P\$ in the N-tensor-factor products of eigen-vectors. One obtains zero otherwise.\hfill $\Box$\\

{\it REMARK 2} The number of combinations of K distinct elements out of a set of N elements is, what is called,  "N over K": \$N!\Big/K!(N-K)!\$ (cf "Combination" Wikipedia).\\

{\it    THEOREM 1. The  BRIDGING THEOREM} reads:   If in complete nondemolition (synonyms: predictive, repeatable)\M of a given event \$P\$ on N single systems, each in the same given state \$\ket{\psi}\$, the \EM result K/N is obtained, and hereby \$N\$ single-system
pure states \$\{\ket{\psi'}_n:n=1,2,\dots ,N\}\$ are produced, and if among
them \$K\$ satisfy \$\ket{\psi'}_n = P_n\ket{\psi'}_n\$ and for
\$(N-K)\$ of them \$\ket{\psi'}_n= P^{\perp}_n\ket{\psi'}_n\$ is valid, then the corresponding ensemble state, in its turn,
satisfies the relations
$$\ket{\Psi'}\N\equiv\ket{\psi'}_1\otimes\ket{
\psi'}_2\otimes\dots\otimes\ket{\psi'}_N=$$  $$
Q\N^K\ket{\Psi'}\N=(K/N)^{-1}
F\N^{P}\ket{\Psi'}\N\eqno{(17)}$$ (except if \$K=0\$ when the last expression should be omitted). In other words, the
\RF operator \$F\N^{P}\$ is exactly completely measured
and the theoretical \RF K/N is obtained.\\

{\it PROOF} It is seen from definition (16) of
\$Q\N^K\$  that all its projector terms, acting on the ensemble state \$\ket{\Psi'}\N\$, give zero except the one term that has its \$P_n\$ factors precisely in the same positions as the \$\ket{\psi'}_n=P_n\ket{\psi'}_n\$ tensor factors in the string \$\ket{\Psi'}\N\$ of single-system final states. This projector term leaves the state unchanged, and hence so does \$Q\N^K\$.\hfill $\Box$\\

In Theorem 1 the {\it \T \RF    EQUALS  the \EX \rf }. Here the basic entity of the
unitary \Q formalism, the \T \rf , 'meets' its
statistical (experimental) counterpart, the \EX \rf .
This establishes the required {\it bridge} from probability to the statistical ensembles in the laboratory.\\

If the reader is not familiar with the concept of nondemolition \m , she (or he) is advised to look it up in the relevant review \cite{FHArxiv}.\\

%4     last (17)
%%%%%%%%%%%%%%%%%%%%%%%%%%%%%%%%%%%%%%%%%%%%%%%%%%%%%%%%%%%%%%%%%%%%%%%%%%%%%
%%%%%%%%%%%%%%%%%%%%%%%%%%%%%%%%%%%%%%%%%%%%%%%%%%%%%%%%%%%%%%%%%%%%%%%%%%%%%
%%%%%%%%%%%%%%%%%%%%%%%%%%%%%%%%%%%%%%%%%%%%%%%%%%%%%%%%%%%%%%%%%%%%%%%%%%%%%
\section{Probabilities of Experimental Relative Frequencises}
%%%%%%%%%%%%%%%%%%%%%%%%%%%%%%%%%%%%%%%%%%%%%%%%%%%%%%%%%%%%%%%%%%%%%%%%%%%%%
%%%%%%%%%%%%%%%%%%%%%%%%%%%%%%%%%%%%%%%%%%%%%%%%%%%%%%%%%%%%%%%%%%%%%%%%%%%%%
%%%%%%%%%%%%%%%%%%%%%%%%%%%%%%%%%%%%%%%%%%%%%%%%%%%%%%%%%%%%%%%%%%%%%%%%%%%%%

We shall consider sequences \$\cS\equiv\{K/N:N=1,2,\dots ,\infty\}\$ of \EX \rf s. In view of the bridging theorem, we can say that they consist of one element from each spectrum of \$\{ F\N^{ P}:N=1,2,\dots ,\infty\}\$. Deeper insight is now required. To this end, let us evaluate the relevant probabilities.\\

{\it THEOREM 2} {\it The Probability-of-Relative-Frequency Theorem} reads: Let \$p\equiv\bra{\psi} P\ket{\psi}\$, where \$\ket{\psi}\$ and \$ P\$ are the given arbitrary single-system state and the given arbitrary projector respectively. We take into account all possibilities \$0\leq p\leq 1\$. Then
$$\bra{\Psi}\N Q\N^K\ket{\Psi}\N=\Big(N!\Big/[K!(N-K)!]\Big)
p^K(1-p)^{(N-K)}\eqno{(18)}$$ is valid,
where \$\ket{\Psi}\N\equiv\prod_{n=1}^{\otimes ,N}\ket{\psi}_n\$ is the ensemble state, \$\ket{\psi}_n,\enskip n=1,2,\dots ,N\$ are all isomorphic to the single-system state \$\ket{\psi}\$, and \$Q\N^K\$ is the eigen-projector of the relative-frequency operator \$ F\N^{P}\$ for \$K=0,1,\dots ,N;\enskip N=1,2,\dots ,\infty\$.
In the expression on the RHS of (18) it is understood that \$p^0=(1-p)^0\equiv 1,\enskip 0\leq p\leq 1\$ and \$0!\equiv 1\$.\\

{\it PROOF} On the LHS(18) we have strings and a sum of strings of single-system entities. Hence, one obtains the corresponding sum of single-system probabilities in each factor separately in each term (cf (16) and the passage beneath it):
$$LHS(18)=\sum_{all\enskip combs_N^K}\bra{\psi}_1
\otimes\bra{\psi}_2\otimes\dots\otimes\bra{\psi}_N
\Big(\prod_{n=1}^{\otimes,N} P^i_n\Big)
\ket{\psi}_1
\otimes\ket{\psi}_2\otimes\dots\otimes\ket{\psi}_N
$$  $$=\sum_{all\enskip combs_N^K}\prod_{n=1}^N\bra{\psi}_n P_n^i\ket{
\psi}_n,\eqno{(19)}$$  where \$i=1,2\$ and \$\hat P^{i=1}\equiv P\$, \$P^{i=2}\equiv P^{\perp}\$. The former projector appears in each term K times, and the latter (N-K) times. Since in each term of the last expression we have multiplication of numbers, irrespectively of which factor contains \$ P_n\$ and which \$P^{\perp}\$, each term equals \$p^K(1-p)^{(N-K)}\$. According to Remark 2, there are "N over K" \$combs_N^K\$.\hfill $\Box$\\

{\it COROLLARY 1} As far as the {\it extreme values} of \$p\$ are concerned, Theorem 2 implies the following simple relations for all K and N values.

(i) if \$p=0\$, then
$$\bra{\Psi}\N Q\N^K\ket{\Psi}\N=\delta_{K,0}\enskip,\eqno{(20)}$$

(ii) if \$p=1\$, then $$\bra{\Psi}\N Q\N^K\ket{\Psi}\N=\delta_{K,N}.\eqno{(21)}$$\\

As it was stated in the Introduction, since the extreme cases belong to the e-e link, they will be discussed elsewhere.

Having clarified the extreme cases, we shall avoid them in our further analysis.\\

{\it COROLLARY 2} Relation (18) shows that for \$0<p<1\$, each eigenvalue \$\{K/N:K=0,1,\dots ,N\}\$ of each relative-frequency operator \$\{ F\N^{ P}:N=1,2,\dots ,\infty\}\$ has a {\it positive probability} in the corresponding $N$-system ensemble state \$\ket{\Psi}\N\$ determined by an arbitrary single-system pure state \$\ket{\psi}\$. Hence, the elements of any sequence \$\cS\$ of \EX \rf s can take any of the relative frequency values with a positive probability.\\

At first glance, it seems hard to reconcile Corollary 2,  with the \RFP (in the Introduction). We will have to return to the weak law and deepen our analysis.\\

%5     last (21)
%%%%%%%%%%%%%%%%%%%%%%%%%%%%%%%%%%%%%%%%%%%%%%%%%%%%%%%%%%%%%%%%%%%%%%%%%%%%%
%%%%%%%%%%%%%%%%%%%%%%%%%%%%%%%%%%%%%%%%%%%%%%%%%%%%%%%%%%%%%%%%%%%%%%%%%%%%%
%%%%%%%%%%%%%%%%%%%%%%%%%%%%%%%%%%%%%%%%%%%%%%%%%%%%%%%%%%%%%%%%%%%%%%%%%%%%%
\section{Quantum Probabilistic Convergence}
%%%%%%%%%%%%%%%%%%%%%%%%%%%%%%%%%%%%%%%%%%%%%%%%%%%%%%%%%%%%%%%%%%%%%%%%%%%%%
%%%%%%%%%%%%%%%%%%%%%%%%%%%%%%%%%%%%%%%%%%%%%%%%%%%%%%%%%%%%%%%%%%%%%%%%%%%%%
%%%%%%%%%%%%%%%%%%%%%%%%%%%%%%%%%%%%%%%%%%%%%%%%%%%%%%%%%%%%%%%%%%%%%%%%%%%%%

We are going to analyze the fine structure of the variance formula
(8), utilizing relation (10), which reads
$$V_N=\sum_{K=0}^N[(K/N)-p]^2\bra{\Psi}\N Q\N ^K\ket{\Psi}\N.\eqno{(22)}$$

We can choose {\it an arbitrarily small positive
number \$\epsilon\$} and consider a possible term, we denote it by  \$\bar K\$, in which
$$[(\bar K/N)-p]^2> \epsilon ^2\eqno{(23)}$$
(an "outside-$\epsilon$" term). Making use of (23),
the fact
that each term in (22) is non-negative, and utilizing the weak law in the form of the basic variance relation (12) in Lemma 3, we can argue as follows.
$$\epsilon ^2\bra{\Psi}\N Q\N ^{\bar K}\ket{\Psi
}\N <
[(\bar K/N)-p]^2\bra{\Psi}\N Q\N ^{\bar K}\ket{\Psi
}\N\leq $$  $$
\sum_{K=0}^N[(K/N)-p]^2\bra{\Psi}\N Q\N ^K\ket{\Psi}\N =
p(1-p)/N,\eqno{(24a)}$$ i. e., we obtain
$$\epsilon ^2\bra{\Psi}\N Q\N ^{\bar K}\ket{\Psi
}\N < p(1-p)/N.\eqno{(24b)}$$

Let us now restrict the single-system probability \$p\$ to a non-extreme  value \$0<p<1\$.

Next, let us take another {\it arbitrarily small
positive number \$\omega\$, independent of
\$\epsilon\$,} and let us define the integer \$N_{\epsilon \omega}\$ so
that
$$p(1-p)/(\epsilon^2\omega )\leq N_{\epsilon \omega}<
[p(1-p)/(\epsilon^2\omega )]+1. \eqno{(25)}$$
Let us consider \$N\$ so large that $$1/N\leq
1/N_{\epsilon ,\omega}\leq\epsilon^2\omega /[p(1-p)].\eqno{(26)}$$ Then, in
view of (24b), for such \$N\$ values (26) implies
$$\bra{\Psi}Q\N^{\bar K}\ket{\Psi }<
[p(1-p)/\epsilon^2]\times
\epsilon^2\omega/[p(1-p)],\eqno{(27a)}$$  i.e.,
$$\bra{\Psi}\N Q\N^{\bar K}\ket{\Psi }\N <\omega. \eqno{(27b)}$$

Therefore, taking into account Corollary 2, (27b), and (25) for the definition of \$N_{\epsilon\omega}\$, the result of our argument reads
$$N\geq N_{\epsilon\omega},\quad\Rightarrow\quad 0<
\bra{\Psi}\N Q\N^{\bar K}\ket{\Psi }\N\enskip
\mathbf{< }\enskip\omega,\eqno{(28)}$$ where \$"\Rightarrow "\$ denotes logical implication.

Hence, the probability of an "outside-$\epsilon$"
\$\bar K\$ term becomes arbitrarily small when
\$N\$ goes to infinity though it is always larger than zero.

We have derived a result that can furnish the desired answer. Let us put it first in
words. Considering an arbitrarily close {\it approximation} (\$p-\epsilon< K/N<p+\epsilon,\enskip\epsilon<<1\$) of the \P \$p\enskip\Big(\equiv\bra{\psi} P\ket{\psi}\Big)\$
 by an \EM result (\EX \rf) \$K/N\$, it turns out that
it can be obtained with a probability
that is arbitrarily close to certainty though it equals 1 for no value of N. To put result (28) with more precision, we can say that we have proved the  following claim, which we present in two equivalent forms.\\

{\it THEOREM 3. The Quantum Probabilistic Convergence Theorem reads: A)} For every \$\epsilon >0\$ and
\$\omega >0\$, if \$N\$ exceeds \$N_{\epsilon\omega}\$ given by (25), the probability of an \EM result (\EX \rf )
\$K/N\$ such that \$|(K/N)-p|<\epsilon\$
is larger than \$(1-\omega )\$, though smaller than 1.

{\it B)} For every \$\epsilon >0\$ and
\$\omega >0\$, if \$N\$ is larger than \$N_{\epsilon\omega}\$ given by (25), the probability of an \EM result (\EX \rf )
\$K/N\$ such that \$|(K/N)-p|>\epsilon\$
is smaller than \$\omega\$, though larger than zero.\\

We can say {\it intuitively} that {\it almost all} elements of the sequence of \EM results \$\cS=\{K_N/N:N=1,2,\dots ,\infty\}\$ are within an arbitrarily small $\epsilon$-neighborhood of \$p\$ in the probabilistic sense. But there is always an, ever smaller,  positive probability for deviation.\\

%6
%%%%%%%%%%%%%%%%%%%%%%%%%%%%%%%%%%%%%%%%%%%%%%%%%%%%%
%%%%%%%%%%%%%%%%%%%%%%%%%%%%%%%%%%%%%%%%%%%%%%%%%%%%%
%%%%%%%%%%%%%%%%%%%%%%%%%%%%%%%%%%%%%%%%%%%%%%%%%%%%%
\section{Concluding Remarks}
%%%%%%%%%%%%%%%%%%%%%%%%%%%%%%%%%%%%%%%%%%%%%%%%%%%%%
%%%%%%%%%%%%%%%%%%%%%%%%%%%%%%%%%%%%%%%%%%%%%%%%%%%%%
%%%%%%%%%%%%%%%%%%%%%%%%%%%%%%%%%%%%%%%%%%%%%%%%%%%%%

{\bf A)} One should note that Theorem 3 actually defines {\it \Q probabilistic convergence}. It is remarkable that, unlike in mathematical convergence, the ensemble \$N_{\epsilon\omega}\$ beginning  with which almost all results are as close as desired is precisely specified (relation (25)).\\

{\bf B)} An alternative formulation of relation (28), i. e., of \Q probabilistic convergence, could be in terms of {\it probabilistic approximations}:

For every \$\epsilon >0\$ and
\$\omega >0\$, if \$N\geq N_{\epsilon\omega}\$, where the latter is  given by (25), the probability that an \EM result (\EX \rf )
\$K/N\$ approximates the measured probability \$p\$ so that \$|(K/N)-p|<\epsilon\$
is larger than \$(1-\omega )\$ but smaller than 1.\\

{\bf C)} Let us compare three kinds of {\it a priori} possible relevant convergences: the mathematical one, the minimally probabilistic one, and the \Q probabilistic one.

In the first, for every \$\epsilon>0\$ there exists an (unspecified) integer \$N_{\epsilon}\$ such that for {\it all} ensembles for which \$N>N_{\epsilon}\$ one would have \$|(K/N)-p|<\epsilon\$. In the second again for every \$\epsilon>0\$ there exists  an (though unspecified) integer \$N_{\epsilon}\$ such that for all ensembles for which \$N>N_{\epsilon}\$ one would have \$|(K/N)-p|<\epsilon\$ {\it with probability one}.

The second is the probabilistic counterpart of the first. But it is weaker in the sense that it cannot guarantee that every individual \M would give a sufficiently close result. Comparing the second with the third, one may be tempted to view them together as special cases of probabilistic convergence.

Such an \$\omega\geq 0\$ approach (remember that \$\omega\$ is the probability of deviation from the desired closeness) necessarily breaks up into its two mentioned special cases. The minimally probabilistic one appears in the case of extreme individual-system (as contrasted to "ensemble")  probabilities \$p\enskip\Big(\equiv\bra{\psi}P\ket{\psi}\Big)\$. It is formally trivial (cf Corollary 1 with relations (20) and (21)). In the case of intermediate probabilities we had to stipulate \$\omega>0\$ (cf relation (25)). Otherwise, our argument would not go through.\\

{\it \textbf{D)})} Returning to a basic initial consideration, one can say that there are two ways to attempt to derive the kind of convergence in probability \M in \qm :  with the help of the {\it
 strong \Q law of large numbers},  or  leaning on the \Q
{\it weak} law of large numbers as it is done in this study,

The \Q strong law of large numbers concerns a
hypothetical infinite ensemble, and it claims that
the corresponding ensemble state vector is a precise
eigen-vector of the frequency operator (acting in the corresponding space) with the eigenvalue \$p\enskip\Big(\equiv\bra{\psi}P\ket{\psi}\Big)\$.
The strong law seems to be more favored
in the literature (see e. g. [5],[7]) though it is
controversial both mathematically \cite{Schack} and
physically.

When I started this study it seemed to me that the strong-law approach is like accessing a Himalayan peak by helicopter, and the weak-law approach appeared to resemble an arduous step-by-step climbing. I would like to explain why I chose the "climbing".

The uncountably infinitely dimensional Hilbert space that is used by the strong-law approach,
is outside the \Q formalism of non-relativistic
\qm , and it is unsupported by empirical evidence in particle physics. Conclusions drawn in this framework may raise suspicion of sufficient physical relevance.\\

Now that I believe that a Himalayan peak has been reached by "climbing", the following idea occurred tto me. One may be reminded of the
{\it paradigm of irrationals} (on the real axis).
These numbers (like \$\pi\$, the ratio between the
circumference and the diameter of the circle) cannot
be precisely expressed, but each of them can be
approximated arbitrarily well by a rational if its
index in a suitable sequence of rationals is large
enough. Contrariwise, the rationals can be given a
precise value in term of operations.

The irrationals and the rationals are {\it essentially different}. It is futile to endeavor to force an irrational into the form of a rational. It is like with the numerous futile {\it perpetuum mobile} efforts in experimental thermodynamics. It may be that \M of the probability of a \Q event has a similar position in nature. It belongs to the weak \Q law; no use forcing it into the strong \Q law.

I would prefer if this idea were wrong and if one could derive \Q probabilistic convergence also from the strong \Q law of large numbers. (Naturally, Loose end 1 below should ne taken into account.) I would prefer this the more so since much serious work with impressive results was done along the lines of the strong \Q law of large numbers. Let me mention, besides references [5] and [7], only two more: \cite{Cassinello} and \cite{Wesep}.\\

{\bf E)} We have derived \Q probabilistic convergence as the answer to the \M of intermediate probabilities. For what is done in the laboratory
\Q statistical convergence, in which there would be  no mention of probability, might be more desirable.  I chose to hang on to the rich language of probability theory in this study. It will be shown elsewhere that the statistical variant can be derived in a natural way.\\

%App
%%%%%%%%%%%%%%%%%%%%%%%%%%%%%%%%%%%%%%%%%%%%%%%%%%%%%
%%%%%%%%%%%%%%%%%%%%%%%%%%%%%%%%%%%%%%%%%%%%%%%%%%%%%
%%%%%%%%%%%%%%%%%%%%%%%%%%%%%%%%%%%%%%%%%%%%%%%%%%%%%
{\it \Large Appeendix}\\
%%%%%%%%%%%%%%%%%%%%%%%%%%%%%%%%%%%%%%%%%%%%%%%%%%%%%
%%%%%%%%%%%%%%%%%%%%%%%%%%%%%%%%%%%%%%%%%%%%%%%%%%%%%
%%%%%%%%%%%%%%%%%%%%%%%%%%%%%%%%%%%%%%%%%%%%%%%%%%%%%

Let us complete the exposition in this article by
'tying up' 2 loose ends.\\

%%%%%%%%%%%%%%%%%%%%%%%%%%%%%%%%%%%%%%%%%%%%%%%%%%%%%

{\it \large LOOSE END 1} It concerns two
sequences of vectors in a given separable complex Hilbert
space that {\it 'meet' asymptotically}.

{\it LEMMA A.1} If two sequences
\$\{\psi_n:n=1,2,\dots ,\infty\}\$ and
\$\{\phi_n:n=1,2,\dots ,\infty\}\$ meet
asymptotically, i. e., if
\$||\psi_n-\phi_n||\rightarrow 0\$ when
\$n\rightarrow\infty\$, and if one of them is
convergent, e. g., \$\lim_{n\rightarrow\infty}\psi_n=\psi_0\$, then also the
other sequence converges to the same element:
\$\lim_{n\rightarrow\infty}\phi_n=\psi_0\$.

In other words, either the two sequences both
converge, and then necessarily to the same limit, or
none of them converges.\\

{\it PROOF} We can argue as follows using the
triangle rule:
$$||\phi_n-\psi_0||=||(\phi_n-\psi_n)+
(\psi_n-\psi_0)||\leq ||\phi_n-\psi_n|| +
||\psi_n-\psi_0||.\eqno{(A.1)}$$ Since the RHS can be arbitrarily
small, so can the LHS. \hfill $\Box$\\

In our case (cf relation (15) and the passage beneath it) the sequence
\$\{F\N^P\Psi\N:N=1,2,\dots ,\infty\}\$ does not
converge in \$\sum_{N=1}^{\oplus,\infty}\cH\N\$ because \$\{p\Psi\N:N=1,2,\dots ,\infty\}\$
does not.

The space \$\sum_{N=1}^{\oplus,\infty}\cH\N\$, as a countably infinite sum, is still a separable Hilbert space with the completeness property of the terms preserved. Hence, each Cauchy sequence converges. Since both mentioned sequences do not converge, they are not Cauchy sequences.

Incidentally, "completeness" simply means that each Cauchy sequence converges. A sequence \$\{\ket{\psi}_k:k=1,2,\dots ,\infty\}\$ is a Cauchy sequence if for every \$\epsilon>0\$, there exist an integer \$K_{\epsilon}\$ such that \$||\ket{\psi}_k-\ket{\psi}_{k'}|| <\epsilon\$ whenever both \$k\$ and \$k'\$ are larger than \$K_{\epsilon}\$.\\

%%%%%%%%%%%%%%%%%%%%%%%%%%%%%%%%%%%%%%%%%%%%%

{\it \large LOOSE END 2}\\

Let us be reminded that in a separable complex Hilbert space (it has finite or a countably infinite dimension) for every projector \$P\$ (event, property) there exist its ortho-complementary projector (opposite event) \$P^{\perp}\enskip\Big(\equiv  I-P\Big)\$. They are always
distinct and \$(P^{\perp})^{\perp}=P\$.\\

{\it LEMMA A.2} For every projector \$P\$ and every norm-one vector \$\ket{\psi}\$ the following equivalences are always valid:
$$\bra{\psi}P\ket{\psi}=1\enskip\Leftrightarrow \enskip\bra{\psi} P^{\perp}\ket{\psi}=0\enskip\Leftrightarrow \enskip P^{\perp}\ket{\psi}=0
\enskip\Leftrightarrow \enskip P\ket{\psi}=
\ket{\psi}.\eqno{(A.2)}$$\\

{\it PROOF First equivalence} Subtracting
\$\bra{\psi}P\ket{\psi}=1\$ from \$\bra{\psi} I\ket{\psi}=1\$,
\$\bra{\psi}P^{\perp}\ket{\psi}=0\$ is obtained.
Arguing in the opposite direction, we can subtract
\$\bra{\psi}P^{\perp}\ket{\psi}=0\$ from
\$\bra{\psi}I\ket{\psi}=1\$, we obtain back the relation \$\bra{\psi}P\ket{\psi}=1\$.

{\it Second equivalence.} On accoint of idempotency of every projector and due to the positive definiteness of the norm in Hilbert space, the second equivalence in more detail reads:
$$\bra{\psi}P^{\perp}\ket{\psi}=0
\enskip\Leftrightarrow \enskip
||P^{\perp}\ket{\psi}||=0
\enskip\Leftrightarrow \enskip
P^{\perp}\ket{\psi}=0.\eqno{(A.3)}$$

{\it Third equivalence.} Subtracting \$P^{\perp}\ket{\psi}=0\$ from \$I\ket{\psi}=
\ket{\psi}\$ the last relation in (A.2) is obtained.
Subtracting the last relation from \$I\ket{\psi}=\ket{\psi}\$, the last-but-one relation is obtained back.\hfill $\Box$\\

\end{document}